\documentclass{article}
\usepackage{authblk}
\usepackage{wrapfig}
\usepackage{calrsfs}
\usepackage{caption}
\usepackage{graphicx}
\usepackage{epsf}
\usepackage{a4}

\begin{document}

\title{The Science and Legacy of Richard Phillips Feynman}
\author[a]{Avinash Dhar}
\author[b]{Apoorva D. Patel}
\author[a]{Spenta R. Wadia}
\affil[a]{International Centre for Theoretical Sciences\\
Tata Institute of Fundamental Research, Bangalore 560089}
\affil[b]{Centre for High Energy Physics, Indian Institute of Science,
Bangalore 560012}
\date{October 2018}
\maketitle

\centerline{\bf Abstract}
\begin{center}
\parbox{15truecm}{This year is the 100th birth anniversary of Richard Phillips Feynman. This article commemorates his scientific contributions and lasting legacy.}
\end{center}

\section{Richard Feynman}
\begin{wrapfigure}{l}{0.25\textwidth}
\includegraphics[width=1.0\linewidth]{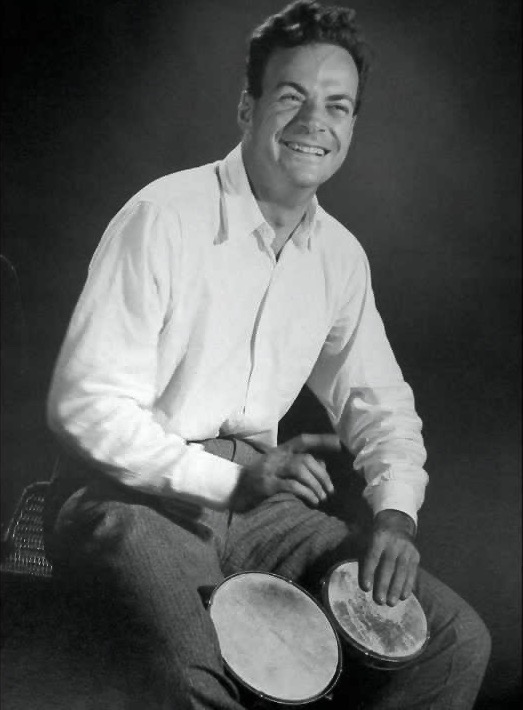} 
\caption*{\it Bongo player and \\ Theoretical Physicist}
\end{wrapfigure}
${\mathcal R}$ichard ${\mathcal P}$hillips ${\mathcal F}$eynman was one of the most influential, iconoclastic and famous physicists of the second half of the twentieth century\footnote{Some general references for Feynman's life and research are Refs.~\cite{RPFwebsite,FeynmanLectures,Feynman100,gleick,mehra,gribbin}.}, who made profound contributions to physics and its visual language (the Feynman diagrams). More importantly, he provided a new meaning to what it means to understand the character of physical law, unfolding it in his uniquely refreshing style to explain the `beauty of world'. 

Feynman was born on 11th May 1918 in New York city, and grew up in Far Rockaway in Queens. He was an undergraduate at MIT during 1935-39, and got his doctorate from Princeton University in 1942, advised by John Wheeler. Besides his work with Wheeler on classical electrodynamics, he wrote a thesis on ``The Principle of Least Action in Quantum Mechanics" that reformulates non-relativistic quantum mechanics in terms of path integrals. During and after his PhD, he was involved with the Manhattan Project till 1945. After that he spent 5 years at Cornell University, where he developed his Nobel prize winning work on how to do calculations in quantum electrodynamics using Feynman diagrams. Feynman moved to Caltech in 1950, and spent the rest of his career there till he died on 15th February 1988. From 1959 onwards he was Richard Chace Tolman Professor of Theoretical Physics at Caltech. During his 38 years at Caltech, he worked on a variety of problems posed by mother nature, and covering all four fundamental interactions. These include the quantum theory of liquid Helium, turbulence, V-A theory of the weak force, quantization of gravity and of Yang-Mills theory, quantum theory of open systems, the parton model of hadronic physics and quantum computing. To each of these problems he made lasting contributions that bear the stamp of his insights and originality. 

He influenced the way physicists think about physics, especially physical processes whose description requires the quantum theory. Feynman's approach to physics was to show how the solution to a problem unravels, aided by a visual language that encapsulates complicated mathematical expressions. 
James Gleick put this very succinctly, ``Feynman's reinvention of quantum mechanics did not so much explain how the world was, or why it was that way, as to tell how to confront the world. It was not knowledge of or knowledge about. {\it It was knowledge how to.}" He went to the heart of the problem he was working on, built up the solutions from simple ground rules in a step by step nuts and bolts way, articulating the steps as he built up the solution, keeping in mind that science is highly constrained by the fact that it is a description of the natural world. He laid bare the strategy of the solution, and was explicit about the various difficulties that need to be surmounted, perhaps now or in the next attempt to solve the problem:
``In physics the truth is rarely perfectly clear." Feynman's attitude to `fundamental physics' is well put in the collection, `The Pleasure of Finding Things Out': ``People say to me, `Are you looking for the ultimate laws of physics?' No, I'm not, I'm just looking to find out more about the world, and if it turns out there is a simple ultimate law which explains everything, so be it, that would be very nice to discover."

His most original contribution is the reformulation of quantum mechanics in terms of `path integrals' and the visual language of Feynman diagrams that goes with it. Today it is the defining formulation of the standard model of elementary particles, non-perturbative gauge theories on a space-time lattice, quantum statistical physics, string theory, quantum gravity etc. Feynman's method and generalizations not only afford a conceptual unity in the treatment of quantum systems and their classical limits, but also give at weak coupling a powerful computational tool using the pictorial representation of processes using Feynman diagrams. They are the daily fare of physicists across the physical sciences. Feynman diagrams are ubiquitous in high energy physics, many body physics and statistical physics.

Feynman was a great teacher: the ``Feynman Lectures on Physics" have inspired generations of physicists from all over the world. They were introductory physics lectures delivered to beginning students of physics at Caltech during 1961-62, but are a learning experience even for mature physicists. Now Caltech has made available their high quality internet edition free \cite{FeynmanLectures}. 
Besides this monumental 3-volume lectures, there are numerous talks and interviews in which we see the essential Feynman emerging. He had deep misgivings about `authority' and was willing to question anybody on any subject that can be discussed using reason. This liberation from authority was especially true when it came to explaining the real world. He said, ``Science is the belief in the ignorance of experts." He took the notion of `understanding' natural phenomena to a different level: ``What I cannot create, I do not understand." 

There was a deliberate lack of `urban polish' about Feynman. His childhood was in a family of very modest means in Far Rockaway, and throughout his life he felt true to his origins and upheld his `boy from the country image'. He is sometimes accused, by the `very cultured', of building up a certain image of himself to be a cult figure. Any historical figure that creates a new culture in a society could be accused of this. History abounds with examples of people like Newton, Einstein, Beethoven, Dylan, Gandhi, all of whom created a new dimension of human culture within their domains. How can one disassociate their persona from what they created? All these people stood for a new way of thinking or doing things. Feynman was one of them. He created a new social culture by making science the centerpiece of any yearning to grapple at the `truth' about the natural world, and went public about it. Not only that, he gave the scientist a new informal look, simplicity of language and flair, so that young boys and girls in far away impoverished lands could also be attracted by the romance, and believe that they can also do physics...the bongo player who also did theoretical physics! And in there you read Feynman talking about all sorts of phenomena...the universe in a glass of wine...the sloshing of turbulent fluids...there is a feeling that there is a sort of unity in science and that there are some laws that bring order to the immense diversity of the natural world.

Feynman was one of the few scientists of his generation to grasp the importance of science in promoting democracy and human rights. He told a Seattle audience \cite{FeynmanLetters} in 1963 that the need of scientists to freely investigate the world while `dealing with doubt and uncertainty...is of very great value, [one that] extends beyond the sciences'. He added, ``I feel a responsibility as a scientist who knows the great value of a satisfactory philosophy of ignorance, and the progress made possible by such a philosophy, progress which is the fruit of freedom of thought...to proclaim the value of this freedom and to teach that doubt is not to be feared, but that it is to be welcomed as the possibility of a new potential for human beings. If you know that you are not sure, you have a chance to improve the situation. I want to demand this freedom for future generations."

Feynman remarked while teaching a course, ``Einstein was a tall man. He could keep his head above the clouds, and his feet on the ground. For those of us who are not as tall, we have to make a choice." His own choice was clearly to keep his feet on the ground. He was often not satisfied with high-level review talks. To find out the details, he would go and talk to the students, technicians and engineers, who did the actual ground work. He brought this attitude, together with his imagination, acumen and honesty, to reveal the simple truth about the explosion of the Challenger shuttle just after take off. He demonstrated \cite{ChallengerYoutube}, as the world watched on television, the influence of cold on the resilience of the O-ring rubber by dropping it into a glass of ice water! Dyson later said \cite{gleick}, ``The public saw with their own eyes how science is done, how a great scientist thinks with his hands, how nature gives a clear answer when a scientist asks a clear question." Feynman himself concluded his personal report saying, ``For a successful technology, reality must take precedence over public relations, for nature cannot be fooled." 

The rest of this article is devoted to giving the reader a tour of the great scientific achievements of Richard Feynman. The tour is roughly divided into three parts: Quantum Electrodynamics and high energy physics, Condensed matter physics and Quantum computing. Also included are some personal reflections of some who knew Feynman to varying degrees, which illustrate his interactions with students. The tour concludes with Feynman's Nobel banquet speech \cite{nobel}. 
 
\section{Space-time approach to Quantum Electrodynamics}
Feynman did his graduate studies in Princeton University under the guidance of John Wheeler with whom he immediately embarked on a quest for a consistent quantum theory of electromagnetism. Feynman had already become familiar with the problems of quantizing Maxwell's electromagnetic theory as an undergraduate student at MIT, from reading the books of Dirac\footnote{Feynman refered to Dirac as ``my hero", and followed his tracks many times while adding his own twists.} and Heitler. During these studies he had learnt that the main difficulties lay with the infinities which one encountered in calculating the self-energy and vacuum polarization effects. The former arises from the familiar classical divergence associated with the interaction of a charged particle with its own field while the latter is a purely quantum effect and arises from the possibility of production of virtual electron-positron pairs. Feynman was also bothered by the infinities due to the zero point energies of an infinite number of degrees of freedom of the electromagnetic field. He sought a new formulation of quantum electrodynamics (QED) that would eliminate these divergences in a consistent manner. He also wanted the new formulation to be manifestly relativistic and gauge invariant.

Feynman's first attempt (with Wheeler) to deal with the divergence due to self-interaction in classical electromagnetism was to eliminate it by definition. The new theory replaced the electromagnetic field by a delayed action-at-a-distance principle and a non-local theory of charges \cite{FeynmanWheeler}. In this theory, each charge interacted with all the other charges but not with itself and the electromagnetic field was a derived, not fundamental, concept. But to get the radiation reaction right, Feynman and Wheeler had to assume that the interactions of moving charges occurred via retarded as well as advanced wave propagation; in fact, to get the magnitude of the radiation reaction right they found that the precise combination involved was half retarded and half advanced. This time-symmetric propagation was to play a crucial role in Feynman's final formulation of QED.

Now although the Wheeler-Feynman theory was able to correctly reproduce all the classical phenomena of Maxwell's theory without having the divergence associated with self-interactions of charges, there were several problems in quantizing it. For example, spontaneous emission of a photon from an atom in empty space was hard to understand in the absence of electromagnetic field degrees of freedom. Moreover, with the measurement of Lamb shift, which was understood as the difference between the self-energies of a free electron and one bound inside an atom, it was clear that the Feynman-Wheeler action-at-a-distance theory could not be the classical starting point for a quantum electrodynamics theory. 

Even though Feynman's final formulation of QED discarded the Feynman-Wheeler approach, this route to it turned out to be crucial for the final formulation in two aspects. One was the concept of half-retarded and half-advanced wave propagation\footnote{Modern relativistic quantum field theory calculations incorporate this through the famous Feynman propagator, which is an essential element of all scattering matrix elements.}, which turned out to be necessary for incorporating positrons in QED\footnote{More generally, it accounts for effects due to antiparticles in any quantum field theory, even in bosonic cases where Dirac's hole theory for antiparticles is not available.} via their interpretation as electrons travelling backward in time \cite{Feynman1949a}, an idea that was earlier advocated by St\"uckelberg \cite{stuckelberg}. The second was the difficulty in quantizing Feynman-Wheeler theory using the known procedures based on Hamiltonian methods. The action for Feynman-Wheeler theory involved entire space-time trajectories, including trajectories for two particles (involving two proper time variables) in the interaction term. So Feynman had to invent a new quantization procedure which would directly use the action or the Lagrangian rather than the Hamiltonian. As described in the next section, this was the motivation to which we owe Feynman's invention of the path integral formulation of quantum mechanics.

Feynman's final formulation of QED was based on the conventional Maxwell theory. It brought back local interactions via the electromagnetic field, and consequently the divergences associated with self-energy and vacuum polarization. The solution to these divergence problems came from an entirely new direction, and involved the new concepts of regularization and renormalization, the first rudimentary steps towards which were already taken in the works of several scientists since early 1930's\footnote{For a history of these efforts, see for example Ref.~\cite{mehra}.}. 

\subsection{Regularization and Renormalization} Regularization is a ``cut-off" procedure by which divergent calculations are rendered finite and well-defined. The divergences are recovered in the limit in which the cut-off parameter is removed. The main difficulty here is that of inventing a relativistic and gauge-invariant cut-off procedure. But once this is done, it enables meaningful calculations in theories like QED. Renormalization is a deeper concept which is based on the fact that because of interactions the measured parameters of mass, charge, etc. are in general different from the corresponding `bare parameters' which appear in the Lagrangian. The main question then is whether all the divergences can be consistently removed by redefining physical parameters in terms of the bare parameters and the cut-off.

Feynman's key contribution to this discussion was to introduce, for the first time, a relativistic cut-off in QED \cite{Feynman1948}. The cut-off had the physical interpretation of the size of an electron. The method reproduced results in agreement with those obtained by Schwinger and others. It even applied to mesons and the theories of strong nuclear interactions. Soon after this, a technically much simpler, relativistic and gauge-invariant regularization procedure was invented by Pauli and Villars, which now bears their names. Feynman quickly adopted this method to his formulation of QED to make the self-energy and vacuum polarization calculations finite. With this ingredient in place, he finally had a simple set of (diagrammatic and visualizable) rules, and an efficient calculational technique, for computing the matrix element of any QED process in various orders of perturbation theory \cite{Feynman1949b}.

The great simplification that Feynman achieved in his way of doing QED was because of his emphasis on the overall space-time point of view, as opposed to the prevailing Hamiltonian point of view. Feynman looked at the whole space-time evolution of a given system at once, rather than tracing this evolution in detail at every instant of time. By forsaking the Hamiltonian method, he was able to accomplish the marriage of relativity and quantum theory most naturally, and also achieve great simplicity in calculations, overcoming many technical and conceptual difficulties.

\subsection{Feynman Diagrams} 
The tools developed by Feynman in the form of space-time diagrams, `Feynman diagrams' as they are
\begin{wrapfigure}{l}{0.35\textwidth}
\includegraphics[width=1.0\linewidth]{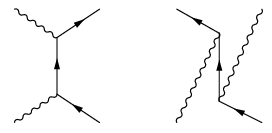} 
\caption*{\it{Tree-level Feynman diagrams for Compton scattering: solid lines represent electron trajectories while wavy lines represent photon trajectories. Time flows upwards and space extends in the horizontal direction.}}
\end{wrapfigure}
called now, were intended to represent physical scattering processes as they happened in space and time. In addition, each element in these diagrams represented a definite mathematical expression. Thus the mathematical expression for the matrix element for any physical process could be written down simply by following the rules for writing mathematical expressions for the various elements in the corresponding Feynman diagram. The illustration shows examples of Feynman diagrams that contribute to the electron-photon scattering matrix element.

Feynman showed that his approach lent enormous simplification to calculations, and made precision tests of QED possible. He had arrived at these rules using his newly developed path integral formulation of quantum theory. As this formulation was not generally known at that time, Feynman's approach took some time to be widely accepted by the scientific community. But once the practical utility of Feynman diagrams and the rules had been demonstrated, they were quickly adopted. Eventually Dyson established complete equivalence of Feynman's approach to the far more complicated approaches of Schwinger and Tomonaga. 

Feynman's space-time diagrams and the associated tools developed by him are truly fundamental, in the sense that they have found applications beyond QED and greatly influenced the development of quantum field theory in general. The diagrams provide a visual representation of the complex microscopic world of elementary particles. Today these tools are an essential element of the scientific training of every theoretical physicist.

\section{Path Integrals}

An early by-product of Feynman's research on electrodynamics was the creation of the `path integral' formulation of quantum mechanics (QM). Early days of QM were dominated by Hamiltonian methods. Both Schr\"odinger's and Heisenberg's formulations of QM, which Dirac showed to be equivalent, used these methods. Feynman could not apply these methods, in any known way, to quantize his
non-standard action-at-a-distance theory for electrodynamics involving space-time trajectories of electrons. He needed a Lagrangian formulation of the subject, so he created one! It happens to be a rare instance, where the subject went from a thesis \cite{Feynman_thesis}, to a review article \cite{pathint_RMP}, to a textbook \cite{Feynman_Hibbs}, with not many publications in between. It also illustrates Feynman's style of work, and the care he took in formulating a new approach.

In an important but not widely known paper published eight years earlier, Dirac had identified the quantity $e^{iS/\hbar}$, where
$S(X(T),x(t))=\int_t^T L(x,\dot{x},t)~dt$ is the action corresponding to the propagation trajectory of a particle from the initial point $x(t)$ to the final point $X(T)$, as the semi classical analogue of the quantum mechanical kernel associated with the time evolution \cite{dirac}.
This was done on the basis of the Hamilton-Jacobi theory and variational methods, and Dirac did not go beyond making this observation since he believed that the correspondence of the action with the phase of the kernel could only work at the semi-classical level, in line with the belief generally prevalent from the beginning of QM.
This correspondence indeed does not work for finite time evolution, but it does when the evolution interval is divided into an infinite number of infinitesimal steps, and a sum over complete set of states is inserted at each time-slice. Feynman asserted that this time-sliced correspondence is actually exact, derived the Schr\"odinger equation from it, and demonstrated how the time-sliced integrals reproduce well-established results of QM.
This definition produced a space-time picture analogous to Huygens's construction of wave propagation. There is an amplitude exp$(iS/\hbar)$ for each possible path a particle could traverse from the initial to the final point, with $S$ being the action for the path. The complete quantum mechanical amplitude for a particle to go from the initial to the final point is the superposition or `sum of all possible paths' between the two points.
Feynman thus established a new general procedure for quantizing any classical system starting from its Lagrangian or action. He published all this in his thesis in the spring of 1942 \cite{Feynman_thesis}.

The meaning of the underlying limiting time-slicing procedure and the precise definition of the `sum of all possible paths' was clarified by Feynman in his famous 1948 review \cite{pathint_RMP}, and several connections to the standard results of QM were added in his authoritative textbook \cite{Feynman_Hibbs}. The power of the path integral formulation of QM lies in the possibility of
decomposing a complex system into several pieces and `integrating out parts of it', which is not at all easy to do in the Schr\"odinger or Heisenberg formulations of QM. Such a partial integration generically converts pure quantum states to mixed states, and has to be implemented on the density matrix and not on the wave-function. It is also unavoidable in dealing with practical problems, because no physical system can be perfectly insulated from its environment and other inaccessible degrees of freedom. Together with his student Vernon, Feynman showed how the behavior of a quantum system of interest can be described in terms of its own variables only, once the external (or environmental) quantum systems that it is coupled to are integrated out. All the external effects are then included in an `influence functional', which depends on the system's own variables, and modifies the undisturbed path integral evolution amplitude $e^{iS/\hbar}$. They explicitly evaluated the influence functionals for the cases of Gaussian
noise as well as linear coupling to environment (modelled as a set of harmonic oscillators), and pointed out how generic fluctuation-dissipation relations emerged from the results. This strategy of tackling a complicated quantum problem by integrating out parts of it, and then describing the remaining system of interest using an `effective action', is now a well-known powerful technique with a large number of applications. 

Since Feynman's original work on QM for non-relativistic spinless particles, the path integral formulation has been extended to include particles with spin, relativity and anomalous quantum field theories (where symmetries of the classical action are broken by quantization). For spin-half particles, it is constructed in terms of anti-commuting Grassmann variables and abstract Grassmann integration, while anomalous theories need careful definition of the integration measure. The quantum path integral amplitude $e^{iS/\hbar}$ can also be converted to the statistical Boltzmann weight $e^{-\beta H}$, by analytically continuing the Minkowski time coordinate to a Euclidean time coordinate. This relation has proved to be extremely useful in carrying over quantum field theory methods to statistical mechanics problems and vice versa. With all these features, the path integral formulation has proved to be quite powerful in dealing with a variety of problems. Indeed, it is routinely usedas a non-perturbative definition of quantum field theories discretized on a lattice. Furthermore, it is essential for quantization of non-abelian gauge theories, which lie at the heart of the Standard Model of strong nuclear and electroweak interactions. Feynman himself used the formulation in diverse applications such as quantum field theories, statistical mechanics models, the polaron problem, superfluidity in liquid helium, Brownian motion and open dissipative systems.

\section{Weak Interactions}
Feynman's interest in this area was kindled at the 1956 Rochester Conference where he began to seriously think about the problem of apparent parity non-conservation in weak interactions. There is the well-publicized incident of Feynman repeating the question of Martin Block (an experimentalist with whom Feynman was sharing accommodation during the conference) in the discussion session, as to whether parity might be violated in K-decays. This was in connection with the hotly debated $\theta-\tau$ puzzle, which referred to the issue of possible existence of two particles which had the same mass, were produced in the same proportion all the time and had all other properties identical, except for having opposite parities. The puzzle would disappear if parity was not conserved in weak interactions, but at that time such thinking was revolutionary. By early 1957, however, this question was settled by Lee and Yang who worked out the consequences of parity violation, and made predictions which were quickly verified by Wu. 

Feynman was excited by these developments as he noticed the appearance of the chiral projection operators $(1\pm\gamma^5)/2$ in these works. They connected with his earlier observations about two-component chiral spinors (whose simplifying features he had already discussed almost a decade earlier), which he believed were more fundamental than the four-component Dirac spinors. Assuming that only one specific of the two two-component chiral projections of all the fermions appeared in weak interactions, he quickly worked out that the law of weak interactions had to be the universal law of Vector-Axial Vector (V-A) interactions. 

Feynman informally reported a preliminary version of this (V-A) theory at the April 1957 Rochester Conference. At that time the available experimental data on neutron decay was considered to be explained by scalar and tensor couplings. Using the (V-A) theory Feynman could not hope to correctly reproduce the experimental observations, and he admitted so in his informal outline of the (V-A) idea at the conference. However, the theoretical beauty of these ideas was compelling and upon re-examining the experiments on neutron decay, it seemed to him that the experimental conclusions were based on rather weak evidence. Convinced about the universality of (V-A) weak interactions, he wrote his definitive paper on the subject (with Gell-Mann) in September of the same year \cite{FeynmanGellMann}. This work emphasized the use of two-component spinors, the current-current form of the weak interactions (possibly mediated by heavy intermediate bosons) and the conserved vector current, all of which were eventually borne out by future experiments.

Feynman had a great sense of achievement in his discovery of the (V-A) law of weak interactions \cite{mehra}. He did not think that the (V-A) equations were quite as beautiful as Dirac's equation or Maxwell's theory, but they filled him with the thrill of a completely new discovery. The (V-A) theory became a starting point for the latter work by Glashow, Salam and Weinberg on the SU$(2)\times$U$(1)$ gauge theory of electro-weak interactions. 

As it transpired, Feynman's discovery was not without controversy concerning his collaboration with Gell-Mann, and priority claims and bitterness, especially on the part of Sudarshan who had also, with Marshak, arrived at the universal (V-A) law by the April 1957 Rochester Conference, though it was not presented there\footnote{An account of this history as reported by Marshak is available in Ref.~\cite{SudarshanMarshak}.}. Feynman tried to make amends and was eventually quite happy to share the credit for the discovery with Sudarshan, Marshak and Gell-Mann.

\section{Parton Model}
In August 1968, while visiting SLAC to see the experimental measurements on deep inelastic scattering of electrons by hadrons, Feynman suddenly realized that these observations were measuring the momentum distribution of partons, the constituent parts of a hadron. Over the previous two decades or so, Feynman had been keenly following the experimental observations and theoretical modelling of composite structure of hadrons, including the models of Fermi and Yang, Sakata and his school, the S-matrix bootstrap hypothesis, etc. He was also aware of the quark model of hadrons of Gell-Mann and Zweig, but in the absence of any experimental observation of fractionally charged quarks, generally physicists did not believe in their existence. Earlier that year in 1968, he had begun thinking about a hadron as a collection of many parts, in a way that was not based on any specific microscopic model of hadrons, but incorporated general features expected of a large class of composite models that could be described by relativistic quantum field theories. The SLAC measurements suddenly seemed to him to bring the partons to life. The SLAC theorist, James Bjorken, who the experimentalists consulted, had already deduced \cite{BjorkenScaling} that the experimental results pointed to scaling in deep inelastic scattering, i.e. he had found that the results depended only on a particular combination of energy and momentum transfer, rather than independently on both the variables, and could be understood on the basis of hard scattering by point-like constituents of nucleons. However, Bjorken used the language and techniques of current algebra, which was rather too abstract for the SLAC experimentalists, who found the partons concept of Feynman, and the associated physical picture of the scattering measuring their momentum distribution, appealing and easy to grasp. So his intuitive interpretation, and the parton model on which it was based, quickly became popular.

Feynman went on to formulate the general problem of deep-inelastic scattering of electrons (equivalently virtual photons) and neutrinos (equivalently virtual massive intermediate vector bosons of the weak interactions) by nucleons in terms of the parton concept, and he extensively lectured on it \cite{partons}. He introduced the concepts of inclusive and exclusive processes---the former averages over all possible outcomes of a scattering process, while the latter describes detailed properties of each possible outcome. In the absence of any specific theory of strong interactions, his ideas were not based on any particular field theory but on the general characteristics of such theories. The partons were to be understood as the basic quanta of the underlying field theory, and the hadronic wavefunction as containing the amplitude for finding configurations of partons with various momenta.

Feynman left the question of whether his partons were merely another name for already proposed quarks, or something completely different, to be answered by experimentalists. As it turned out, partons could include the yet to be introduced gluons as well. With the discovery of asymptotic freedom and the proposal of describing nuclear forces by an underlying gauge field theory called Quantum Chromodynamics (QCD), partons came to be increasingly identified with quarks and gluons, and that clarified several aspects of the parton model\footnote{For example, partons were assumed to behave as almost free particles at high momentum transfers, yet only normal hadrons were seen in the debris of every deep inelastic scattering process. Normally, for a collection of almost free particles, one might have expected a parton to shoot out once in while. In QCD, because of asymptotic freedom and confinement, quarks inside a hadron behave as almost free particles at large momentum transfers, but eventually undergo a series of soft momentum transfer processes with other quarks to finally emerge as part of final state hadrons.}. Today physicists have a more sophisticated understanding of partons as the constituents of hadrons---experimental observations are consistent with only half of the momentum of a hadron being carried by the constituent quarks (and quark-antiquark pairs from the sea); the other half is carried by the gluons, the gauge field quanta of QCD. In later years, Feynman increasingly used the language of quarks and gluons for hard processes, such as quark and gluon jet production, a phenomenon that was predicted by him (with Rick Field \cite{jets}). In his words \cite{QCDjets}, ``the more we were guided by the principles of QCD, the better the things fitted, and the better were the predictions." Here at last was evidence enough for him that nature required the identification of partons with quarks and gluons! 

\section{Quantization of Einstein Gravity and Yang-Mills theory}
In the mid-1950's and through the 1960's, Feynman undertook investigation of the quantization of Einstein's theory, because there ought to be a consistent quantum theory of gravity even though there would be no practical application given the weak strength of the gravitation field. ``My interest in it is primarily in the relation of one part of nature to another." \cite{Gr_ghosts} His approach was to begin with a massless spin-2 field, the graviton, described by a symmetric tensor field coupled to the energy-momentum tensor of a scalar field, in flat Minkowski space-time. Gauge invariance is manifest in the fact that the graviton has only two polarizations as opposed to the five independent spatial components that describe a spin-2 field. He succeeded in showing that a consistent theory naturally includes gauge invariance, and leads to Einstein's theory. (Here by gauge invariance we mean general coordinate invariance). This approach to General Relativity is discussed in his `Lectures on Gravitation' \cite{Gr_PresKip}.

Feynman then embarked on the task of `quantizing' this theory using the method of Feynman diagrams, given that Newton's constant and hence gravitational effects are very small. He first found that at the one loop level there was something wrong if one went about calculating with the rules that were suggested by QED. The covariant gauge answer was not unitary! This happened for not only Einstein's gravitational theory but also for the simpler Yang-Mills theory. In a sense this happened because covariant gauge fixing does not eliminate the time-like components of the graviton and the time-like component of the gluon.

He then invented a method of doing a gauge invariant unitary calculation by expressing a closed loop diagram in terms of on-shell tree diagrams \cite{Gr_loop-tree}. Accepting this as the correct way to obtain a gauge invariant and unitary answer, he found that the difference with the closed loop Feynman diagram in gravity can be compensated by a `ghost' particle which has spin 1 but Fermi statistics. He did the same but simpler calculation for the Yang-Mills theory and discovered the same phenomenon, except that the ghost particle has spin zero \cite{Gr_ghosts}. This was a startling discovery that would steer the course of the subject. 

This was subsequently extended to all higher loops by Bryce DeWitt \cite{Gr_DeWitt}, and by Faddeev and Popov \cite{Gr_FaddeevPopov}, using the Feynman path integral rather than Feynman diagrams. In a gauge theory, the Faddeev-Popov method is manifestly unitary, and the path integral measure restricted to the covariant gauge fixing condition, and naturally leads to a non-trivial determinant. The ghost particles of Feynman emerged as auxiliary fields in exponentiating the determinant. It should be mentioned that this discussion is restricted to field configurations near the orbit of the gauge field $A_{\mu}=0$. As one explores large field configurations `simple gauge fixing' is not possible in a non-abelian gauge theory \cite{gaugefixing}. The Lagrangian of the gauge theory, plus the gauge fixing term, plus the ghosts, exhibits a new symmetry called the BRST (Becchi-Rouet-Stora-Tyutin) invariance, and the correct principle for covariant quantization is BRST invariance. This symmetry has many interesting applications, including to string theory \cite{Gr_Polchinski}.

Yang-Mills theory is a well defined theory at all length scales. At weak coupling, the theory was shown to be asymptotically free by Gross and Wilczek, and by Politzer \cite{Gr_gross}. An exact non-perturbative formulation was given by Wilson \cite{Wilson}, on a hypercubic Euclidean lattice where the degrees of freedom are the group elements on the lattice links (and not gauge fields valued in the Lie algebra). A continuum limit consistent with asymptotic freedom exists. The situation with the quantization of Einstein's theory is far more complex. Unlike the case of Yang-Mills theory, where the original classical formulation in terms of a non-Abelian gauge field is minimally modified to define a precise relativistic quantum field theory, there is no obvious way to define a quantum theory of gravity. The diagramatic method which Feynman pioneered fails to give sensible answers. Neither is there a space-time lattice formulation in 4-dim which leads to a sensible quantum theory of gravity. 
The only known formulation which has well defined high-energy (short-distance) properties, and which reproduces Einstein's gravity at low energies (long distances), is string theory. One of the main consequences of this ultra-violet completion of the theory of gravity is that the fundamental degrees of freedom of `quantum gravity' are not only the gravitons of Einstein's theory, but other `stuff' called branes as well \cite{Gr_Polchinski}.

\section{Quantum Theory of Liquid Helium and Superfluidity}
Feynman moved to Caltech from Cornell in 1950. During his first decade at Caltech, he devoted considerable time to several outstanding problems in condensed matter physics, including superfluidity in liquid helium, superconductivity and the polaron problem. We summarize here his most significant contribution to condensed matter physics, which had to do with (i) understanding superfluidity of Helium-4 and the Landau theory from a microscopic quantum mechanical point of view; (ii) the presence of quantized vortices in Helium-II, and (iii) quantum turbulence. Feynman's approach to the spectrum and excitations of Helium-II is to discover the solution of the theory in close contact with experiment, in a sense the wavefunction describing phonons, rotons and vortices is being built up by a theory-experiment dialogue.

Neutral Helium gas under normal pressure liquifies at a temperature of 4.22K. At a critical temperature of $T_c=2.176K$ it undergoes a superfluid phase transition, called the $\lambda$ transition due to the shape of the specific heat curve as a function of temperature. The two phases are named Helium-I and Helium-II, and Helium-II exhibits superfluidity. A phenomenological model to explain experimental facts about Helium-II, the 2-fluid model, was put forward by Landau and also independently by Tisza, while London advocated a Bose-Einstein condensation at $T_c$. Landau went ahead and formulated 2-fluid hydrodynamics, which was successful in explaining many features of Helium-II. In the 2-fluid model, one component is the superfluid, and the other component is the normal fluid comprising of a gas of phonons and rotons. 

Feynman undertook to understand and work out a first principles understanding of superfluidity in the strongly interacting system of Helium atoms. He wrote 6 papers on this subject during 1953-57 by himself, and three with his student Michael Cohen. He also wrote an excellent exposition of the subject \cite{He_QM-LH}.

\subsection{The $\lambda$ point and below}
His first paper in the subject was on `The Atomic Theory of the $\lambda$ Transition in Helium' \cite {He_lambda-pt}. He formulated the partition function of Helium as a path integral for a system of pairwise interacting bosons. The path integral was in imaginary time and described the partition function. Now knowing the characteristics of the Helium atom, he approximated the interatomic interaction as having a hard core, and argued that the effect of the interaction would be to renormalize the mass of the Helium atom. The path integral then reduced to that of an almost `non-interacting' system of particles governed by Bose-Einstein statistics. In his own words, ``The physical idea which plays a central role is that in a quantum mechanical Bose liquid the atoms behave in some respects like free particles." The approximate partition function which is a sum over the cycles of the permutation group, maps onto a classical polymer problem. Feynman argued that near the transition point the partition function is dominated by large cycles, which have a large entropy that can equal and exceed the energy causing a transition. The mechanism is similar to the Kosterlitz-Thouless-Berezinsky transition in the 2-dim XY model. With the approximations Feynman used, he obtained a transition temperature which was roughly correct, but the phase transition was of third order. Subsequent more accurate numerical work (decades later) implemented Feynman's path integral program (without making the approximations Feynman had made), to give a quantitative theory of the Helium transition and other properties \cite{He_Ceperley}.

A couple of years later in a paper titled `Application of Quantum Mechanics to Liquid Helium' \cite {He_QM-LH}, he gave a physical picture of the phase transition in terms of the proliferation of a few large quantized vortices: ``Then suddenly the rings of very largest length are of importance. The state with one vortex line (or very few) which winds and winds throughout the liquid like a near approximation to a Jordan curve is no longer of negligible weight. The superfluid is pierced through and through by the vortex line. We are describing the disorder of Helium-I. At first the curve doesn't make full use of all its orientations and higher entropy. But as the temperature rises a little more it squeezes into the last corners and pockets of superfluid until it has no more flexibility available. The specific heat curve drops off from the transition to a smooth curve, and the memory of the possibility that Helium can exhibit quantum properties in a unique way is lost in the profusion of states and in disorder, as it is for more usual liquids." Incidentally, in the same paper Feynman discussed quantum turbulence for the first time, in terms of quantized vortices. 

The Landau 2-fluid theory has no mention of Bose-Einstein statistics, and was based on the assumption that there are no other low energy excitations in Helium-II besides density waves (phonons). Feynman recognized this as a fundamental property of the system, which would explain the mysterious properties of Helium-II, and set out to understand this from first principles \cite{He_1}. He was also aware that the transition temperature, and the specific heat curves of Bose-Einstein condensation of a system of free bosons, are significantly different from those of Helium. All his works on Helium, except the one on the $\lambda$ transition, used the Schr\"odinger picture (not path integrals), and based the analysis of the ground state wavefunction and its excitations on physical arguments involving the variational method.

\begin{wrapfigure}{l}{0.25\textwidth}
\includegraphics[width=1.0\linewidth]{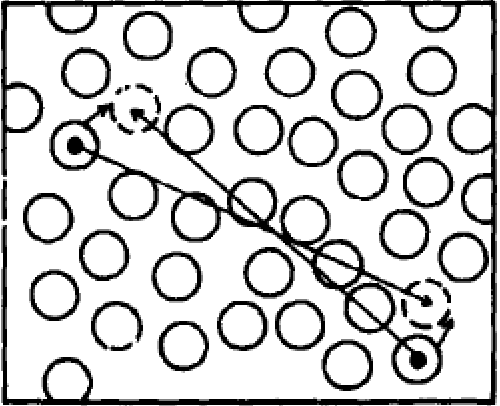} 
\caption*{\it{Two configurations (solid and dotted) resulting from large displacements (long arrows) of atoms can actually be accomplished by much smaller adjustments (short arrows) because of the identity of atoms.}}
\end{wrapfigure}

Using elementary reasoning, Feynman argued that the ground state wavefunction of liquid Helium is non-degenerate, real positive and without nodes. Given the repulsive nature of the interatomic interaction of Helium, the wavefunction vanishes if the atoms come within a distance of $2.14\AA$. Since the Helium atom has spin zero, the wavefunction is totally symmetric in the positions of all the atoms. He therefore argued that the low energy spectrum consists entirely of density waves (phonons) with a linear dispersion law. No other long wavelength (low energy) excitations are allowed by Bose-Einstein statistics. Any other move of Helium atoms over long distances would be a permutation of the atoms that would leave the wavefunction unchanged! See illustration from Ref.~\cite {He_QM-LH}. This was the key insight. 

The excitation spectrum as argued by Landau from a hydrodynamic point of view consists of phonons and rotons. The latter are higher mass excitations with the wave number equal to inverse inter-atomic spacing. To calculate this spectrum, Feynman set up a variational calculation to derive the Landau curve. He obtained a formula in terms of the liquid structure factor (2-point function of the density operator in the ground state), which could explain the shape of the Landau curve qualitatively but not quantitatively.

To get better quantitative agreement, Feynman and his student Michael Cohen devised a new trial wavefunction that took into account the `back flow' of the fluid around the rotons. Feynman's picture of a roton was that of a ring of Helium atoms so small that only a single atom could go through it. While the original Feynman wavefunction contained only one quasi-particle operator, the Feynman-Cohen wavefunction contained two quasi-particle operators. The Feynman-Cohen theory \cite{He_Cohen} considerably improved the agreement with experiment \cite{He_Pines}. Perhaps, in order to do better, one will need to include a higher polynomial of quasi-particle operators\footnote{We thank H. R. Krishnamurthy for a discussion of this point.}.

\subsection{Quantized vortices and quantum turbulence} 
To incorporate local slow velocity variations of the atoms, Feynman introduced local phases $S({\bf r})$ in the wavefunction, which can account for the velocity $v({\bf r}) = \frac{\hbar}{m}\nabla S({\bf r})$. It turns out that liquid Helium-II does have vortex excitations as conjectured by Onsager, and discovered independently by Feynman. He went on to give a qualitative picture of why and how the vortices can lead to a turbulent flow. This effect was observed shortly afterwards by Viven \cite{Viven}. Vortices interacting with each other, and with other excitations namely rotons, play a fundamental role in the description of turbulent flow in Helium. This is a very current and fascinating subject in quantum fluid dynamics \cite{He_Viven}.

\section{Computation}
Feynman had a life-long interest in mathematical puzzles, tricks
that speeded up complicated calculations as well as novel ways to carry
out integrations. This interest naturally expanded to cover design of
algorithms and efficient use of computers. While still a graduate student
at the Princeton university, Feynman was recruited to work on the Manhattan
project at Los Alamos. One of his assignments there was to look after the
`IBM group', calculating the energy release for different designs of the
plutonium implosion bomb. They used decks of program cards to carry out
the calculations on the IBM machines, but the speed was much slower than
required. Feynman dramatically boosted this performance, figuring out
strategies of parallel computation, and the problem was solved in time.
Several independent computational threads were executed together, by
interspersing different sets of cards with a phase lag. Also, with proper
organization, a change in a parameter (or a mistake in the calculation)
at some step in the calculation affected only nearby steps; so only those
nearby steps needed reevaluation instead of the whole calculation. Feynman
also constructed a fast iterative method to calculate logarithms bit by
bit, which was implemented years later on the Connection Machine.

In the 1950's, rapid advances in molecular biology prompted Feynman to think
about manipulating and controlling things on a tiny scale. His inspirational
talk ``There's plenty of room at the bottom" was an open invitation to enter
a new field of physics \cite{feyn_nano}. He pointed out that the entire
Encyclopedia Britannica can be written on the head of a pin, by reducing
the size of writing by a factor of 25000, and a little dot would still
contain 1000 atoms in its area. He argued that improved photo- and
ion-lithography techniques would be able to do that, and so pack enormous
amount of information in a small space. He emphasized that there are no
hurdles of principle in reducing the resolution of an electron microscope,
or the size of mechanical devices and computer circuits, all the way to the
atomic scale. He stressed that the tiny machines would not just be a scaled
down version of the macroscopic ones, but would require redesign in
accordance with physical laws, and he held that out as a challenge.
To motivate research in this field, which is now called nanotechnology,
he personally offered two prizes of \$1000---one to the first person who
prints the page of a book with the linear scale reduced by a factor of
25000, and the other to the first person who makes an operating electric
motor that is only 1/64 inch cube. The second prize was awarded within a
year to Bill McLellan, and Feynman was slightly disappointed that no
major new technique was developed for it. The first prize awarded a
quarter of a century later to Tom Newman, who accomplished it using
electron beam lithography.

Feynman loved to get involved in the nitty-gritty details of the projects
that he liked. To be close to his son Carl, who became a computer scientist,
Feynman worked as a consultant for the Thinking Machines Corporation in the
summer of 1983. They developed the Connection Machine, which was a massively
parallel computer working in the single-instruction-multiple-data mode.
The machine was designed as a 20-dimensional hypercube, and Feynman was
assigned the job to design the router that would allow the processors to
communicate with each other. To avoid traffic-jams, the messages needed to
be held in buffers until the required path became free. Feynman converted
the message passing problem to a set of partial differential equations,
and came up with a solution that required five buffers per chip. That was
better than the seven buffers per chip solution obtained by the engineers
using conventional discrete analysis. It had to be ultimately used because
the small chips could accommodate only five buffers, and it worked.

In his later years, Feynman became interested in the potentialities and
the limitations of computers, as determined by the laws of physics,
specifically quantum physics. In 1982, he taught a course at Caltech titled
`The Physics of Computation', together with John Hopfield (a biologist)
and Carver Mead (a computer scientist). The syllabus of that course was
vague, and the lecturers covered various topics in a rather chaotic manner,
often without knowing what would come next. At times, Feynman talked about
what he had thought of the night before while falling asleep. Three
different view-points, and open critique of each other's analysis, made
a thrilling combination, and many students religiously attended the course.
In this environment, the course material rapidly grew in content, and one
course subsequently split into three courses. What Feynman talked about
developed into `Quantum Computation', what Hopfield discussed became
`Neural Networks', and what Mead talked about led to `Neuromorphic
Systems'.

After the split, Feynman titled his course as `Potentialities and
Limitations of Computing Machines'. He took help of several guest
lecturers to introduce various topics. Then in his inimitable style,
proceeded to develop the subject from his own perspective, without
worrying about what all was already known. A refined version of what was
taught in that course, and the exciting ideas that developed from it,
is now available as two excellent books \cite{feyncomp1,feyncomp2}.
In particular, the concept of quantum computers developed there has now
become a thriving field of research.

Feynman's motivation in pursuing the subject of quantum computation was
two-fold. The first was the realization that with shrinking size of its
elementary components, sooner or later, the computer technology will
inevitably encounter the dynamics of the atomic scale. The laws that
apply there are those of quantum mechanics and not those of electrical
circuits. The classical computational framework based on Boolean algebra
is inadequate in the quantum regime. Some aspects have to be generalized,
while some others need reformulation. Quantum objects possess both
`particle-like' and `wave-like' properties---the discrete eigenstates
that form the Hilbert space basis as well as the superposition principle
that allows for simultaneous existence of multiple components. In what
way would this combination alter the axioms of the classical information
theory, and in what way would that change the design and the performance
of computers? Feynman wanted to explore that.

The second motivation was that quantum computers would be ideal for
simulation of physical systems. In Feynman's own words, ``Nature isn't
classical, dammit, and if you want to make a simulation of nature, you'd
better make it quantum mechanical, and by golly it's a wonderful problem,
because it doesn't look easy." Feynman was aware that classical numerical
investigations of many body quantum systems and quantum field theories,
using importance sampling methods to sum over contributions of various
allowed configurations (e.g. Monte Carlo calculations in lattice QCD),
are highly inefficient. He knew that a quantum computer would sum up many
such contributions in one go, as superposition in his path integral
formulation of quantum dynamics, and produce much better results. He
wanted to realize this advantage, although he did not quantify it in terms
of what is called computational complexity. Another aspect that he did not
deal with was imperfections in the physical device components, and what
would be required to make a quantum computer fault tolerant. A decade
passed before computer scientists came up with answers to these questions,
that the quantum advantage in computational complexity can be exponential
in the input size, and that error correction procedures can allow the
computer to work for as long as one wishes provided that the error rate
is below a specific threshold, and that is when research in the subject
really took off.

Feynman looked up on the subject of computation as engineering. The laws
of physics are well-established, and the task is all about converting the
known science to useful technology, i.e. make something to do something.
He wanted to figure out the limits the laws of physics placed on
computation---what we can and cannot do with computers and why. He was
also perplexed by the exponential growth of computational resources in
tackling hard problems on present-day computers, and wondered whether
human intelligence can get around that. In his lectures, he described the
computer as a `file clerk', who most of the time shuffles data from one
place to another and does some processing of information intermittently.
This analogy is suitable for explaining many concepts: computer organization,
instructions, hierarchy of languages, finite state machines as well as
universal Turing machines. In discussing formal computer science topics
such as computability, information theory and coding, he skipped many
details but firmly emphasized the essentials needed in practice. He
illustrated reversible computation and constraints of thermodynamics of
computing, by constructing and working out simple examples in his own style.
His original contribution was of course about designing quantum computers.
He explained how quantum interactions can implement reversible logic
operations, constructed a Hamiltonian that would execute a set of
instructions step by step, and pointed out that quantum computers will
have novel capabilities because certain quantum correlations cannot be
simulated by a local probabilistic classical computer.

Feynman's course was largely treated as curiosity, and no one came
forward to teach it after he taught it for several years. Still he
was very keen that his lectures on computation be published, just
like several of the courses that he taught at Caltech were. Tony Hey did
the painstaking job superbly, guided by Feynman's notebooks and tapes of the
lectures. The two books that came out posthumously \cite{feyncomp1,feyncomp2}
are now pioneering contributions to a field with tremendous technological
potential.

\subsection*{Acknowledgments}
We would like to thank Dipan Ghosh and Swaminathan Kailas for inviting this article for Physics News. We thank Rajan Gupta for providing a copy of his homework assignment corrected by Feynman, and Sandip Trivedi for sharing his reminiscences of Feynman. The images of Feynman that we have used in this article are taken from various internet resources. SRW would like to acknowledge the Infosys Foundation Homi Bhabha Chair at ICTS-TIFR.

\newpage
\noindent\fbox{\parbox{0.98\linewidth}{
\begin{center}
{\large\bf My Fond Memories of Feynman as a Teacher}
\end{center}
 
I arrived at Caltech, in Fall 1980 as a graduate student in High Energy
Physics. As I familiarized myself with the new environment, I learned many
stories about Richard Feynman, not all of them charitable, and how the Caltech
community absolutely adored him. He was suffering from cancer at that time,
but the medical treatment he received kept him going. The following story
of my interaction with him illustrates his persistent hunt for innovative
solutions to physics problems, as well as his irrepressible curiosity and
generosity.

\hspace{0.5truecm}
I took the course `Topics in Theoretical Physics' taught by Feynman in
Spring 1983. Feynman discussed his ideas about QCD in that course. In the
later part of this course, Feynman decided to talk about what would happen
to QCD at finite temperature. He was using a flux tube picture to describe
confinement in QCD \cite{feyn_YM}, and argued that the fluctuations of the
flux tube with increasing temperature would increase the entropy and produce
a transition to the deconfined phase. This was in analogy to his work on
behavior of vortices in liquid He, which I was not aware of. I was working
on lattice gauge theories at that time, and pointed out to him that the
situation for SU(2) and SU(3) gauge theories was quite different. The
deconfinement transition is of second order for SU(2), but of first order
for SU(3). The difference is due to the the center symmetries Z(2) and Z(3)
of the gauge groups, which control the finite temperature dynamics of the
Wilson line, and Monte Carlo calculations have confirmed it. Feynman wanted
a physical picture, and not a lattice QCD description in terms of the
abstract Wilson line. I argued that a description using the flux tube
connecting a quark and an antiquark cannot be the full story. The baryons
would have to play a role, even though they were too heavy compared to the
transition temperature. Our arguments did not converge for a few lectures,
because we could not adapt them to each-other's language. After one such
inconclusive discussion in the lecture, I left for lunch. When I had gone
just outside the Lauritsen laboratory building, a student informed me that
Feynman wanted to tell me something. So I returned to the building, entering
at the bottom of the stairwell. Feynman was waiting at the top. He shouted,
frantically waving his arms, ``Percolation, percolation!", and went back to
his office. I was too stunned to say anything. I met him later in the day,
and he asked me to make a model of this picture. I did that, explaining how
the dynamics of the same Z(3) center symmetry governs the baryons as well
as the Wilson line. Feynman was happy with this result, and his secretary
(Helen Tuck) gave me the privilege of calling him at his home to discuss
physics. I prepared a draft of the paper, and asked Feynman about including
him as an author. He declined, but also said that he would proof-read the
paper. He carefully went through the draft, made some changes in what I had
written, and then the paper was published \cite{patel_flux}.

\hspace{0.5truecm}
Closer to my graduation, I asked Feynman whether he would write a letter of
recommendation for me. He said ``with pleasure", and that was good enough
for me. He also became my thesis examiner. My thesis was on `Monte Carlo
Renormalization Group for Lattice QCD'. He was not impressed by the brute
force numerical simulations using a lot of computational power, and used to
remark with a chuckle ``make a quantum computer and all your problems will
be solved".

\begin{wrapfigure}{l}{0.35\textwidth}
\includegraphics[width=1.0\linewidth]{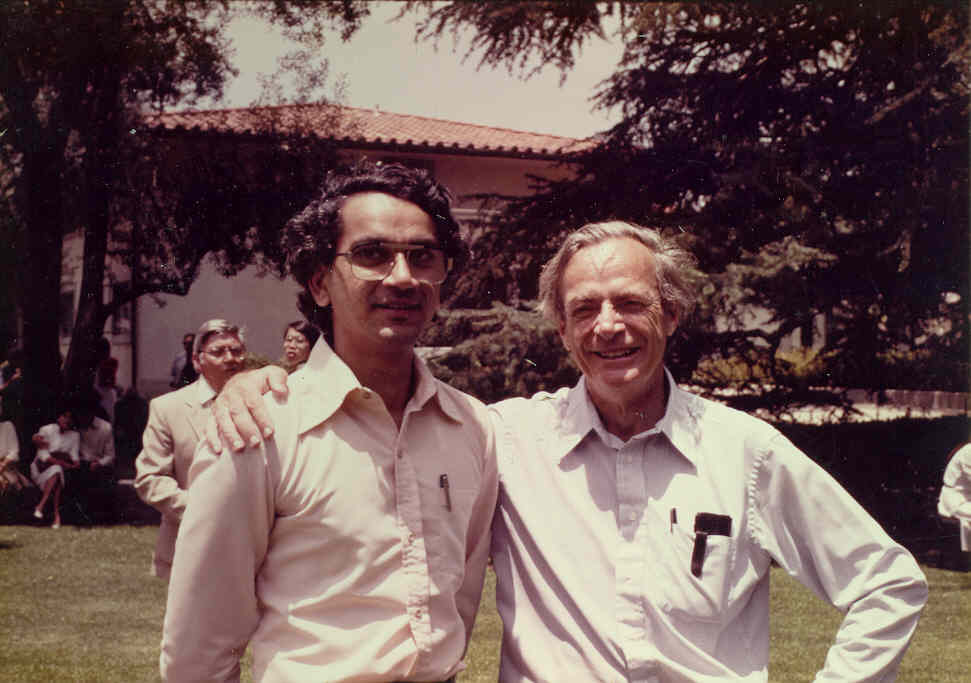}
\caption*{\it{Apoorva and Feynman}}
\end{wrapfigure} 

\hspace{0.5truecm}
Feynman often attended the convocation ceremony, and he was present on
my convocation too. I was a member of the Caltech Men's Glee Club, which 
was to perform at the beginning of the ceremony. On getting the cue, I
strode right down the middle of the congregation to the back of the stage,
sang ``Hallelujah", and then returned to my seat. I met Feynman after the
ceremony. He asked me whether my name had any meaning. I said, ``Yes, it is
an adjective. Can you guess?" He immediately replied, ``Related to what,
truth, beauty?" I said no, it means ``unprecedented". He smiled, and I had
my photograph taken with him. After leaving Caltech, I could not interact
with him again, unfortunately. I learned about his passing away, when I
was working at CERN.

{\sl\hfill ---Apoorva D. Patel}}}

\newpage
\noindent\fbox{\parbox{0.98\linewidth}
{\begin{center}
{\large\bf A Conversation with Feynman}
\end{center}

\begin{wrapfigure}{l}{0.25\textwidth}
\includegraphics[width=1.0\linewidth]{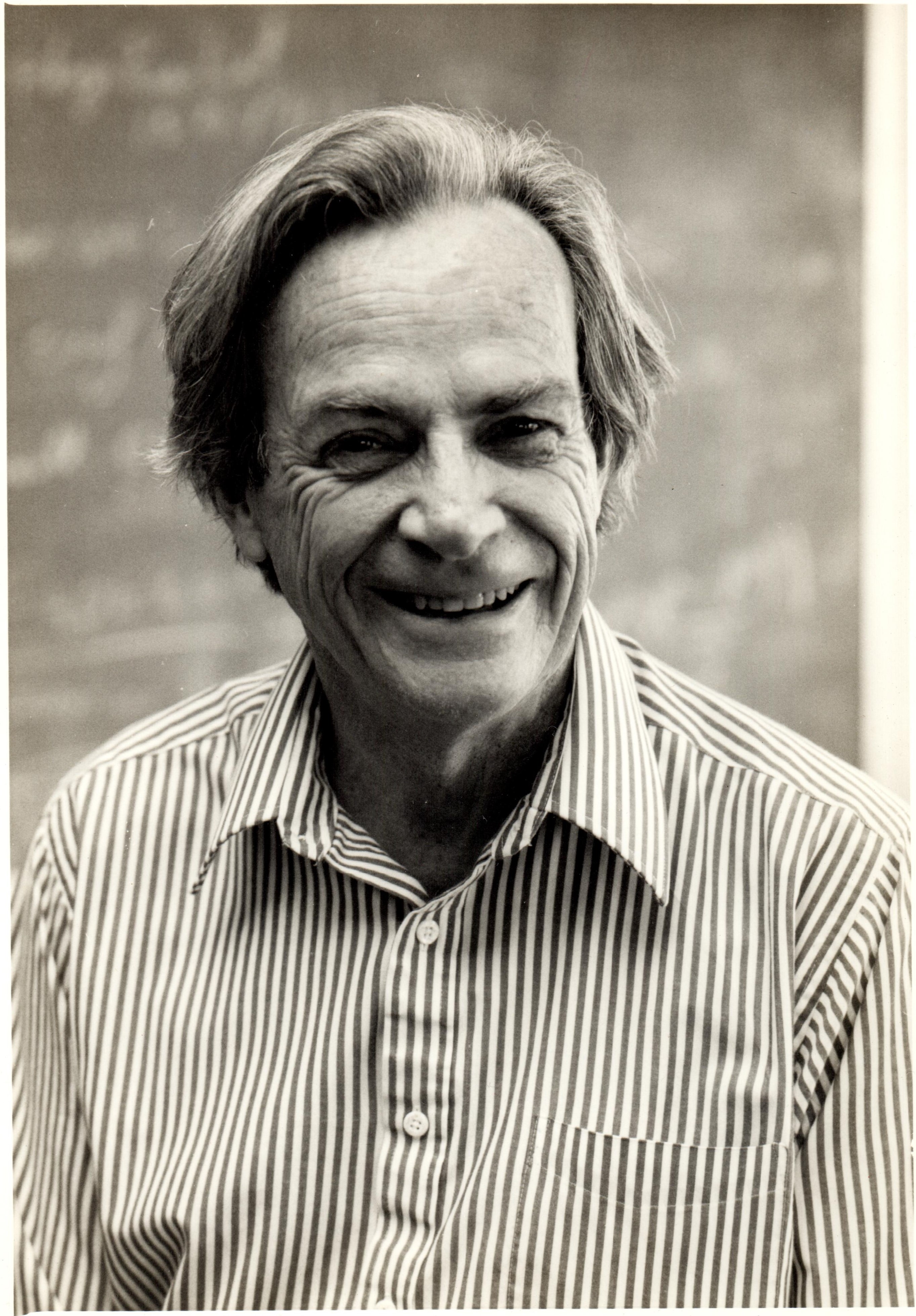} 
\caption*{Feynman in 1986}
\end{wrapfigure} 

In 1986, I visited Caltech for a couple of weeks because my friend and colleague Sumit Das was a post-doctoral fellow there. The memory of my meeting with Feynman is still very vivid. I was keen to speak with Feynman, so I kept pacing outside his office till a surge of courage took me right in. The office was not well-lit, and the furniture was of a darkish hue. He was sitting there in his casual style, and was very welcoming when I asked to speak with him. My reason was that I became a physicist because of his Lectures on Physics! He called me in and had me sit near him, and asked, ``What are you working on?" I replied, ``String theory." He asked me immediately, almost as a reaction, ``What are the important problems of particle physics?" My answer included calculation of the parameters of the Standard Model that would explain the mass hierarchy of leptons; the smallness of CP violation etc.. He was satisfied, and then he wanted to know about some interesting result in string theory. I was surprised, especially because Caltech had distinguished string theorists, but decided to tell him about Dan Friedan's derivation of Einstein's equations as the vanishing of the beta function using the sigma model approach. I started out by saying, ``Let us consider the Feynman path integral," and he snapped back laughing, ``O, what is that?" Continuing, I presented a short derivation of the beta function equation. He was very impressed by this connection: ``You mean renormalization stops!, and gives Einstein's equations."
 
\hspace{0.5truecm}
He mentioned that he had his worries about string theory, but then he had been wrong on many occasions in his career, including electro-weak unification. So I should not be discouraged. He then started to explain to me his work on the existence of a mass gap in 2+1 dim pure Yang-Mills theory, using large gauge transformations and a variational method similar to the one he had used for explaining superfluidity in liquid Helium. He drew an analogy between large gauge transformations in the gauge theory and Bose statistics in liquid Helium \cite{feyn_YM}. This was a truly original way to approach the problem. It was a wonderful discussion because my work with Sumit Das from a few years ago had applied the self-consistent method of instantons in the Euclidean path integral to the same problem \cite{daswadia}. After the discussion he remarked, ``The 4-dim case is too difficult, because of the goddam logarithms!, and that time is running out for me." I did not quite understand the import of the last statement at that time. Meeting Feynman was a truly very intense and inspiring experience.

{\sl\hfill ---Spenta R. Wadia}}}

\begin{figure}
\begin{center}
\fbox{\includegraphics[width=0.6\linewidth]{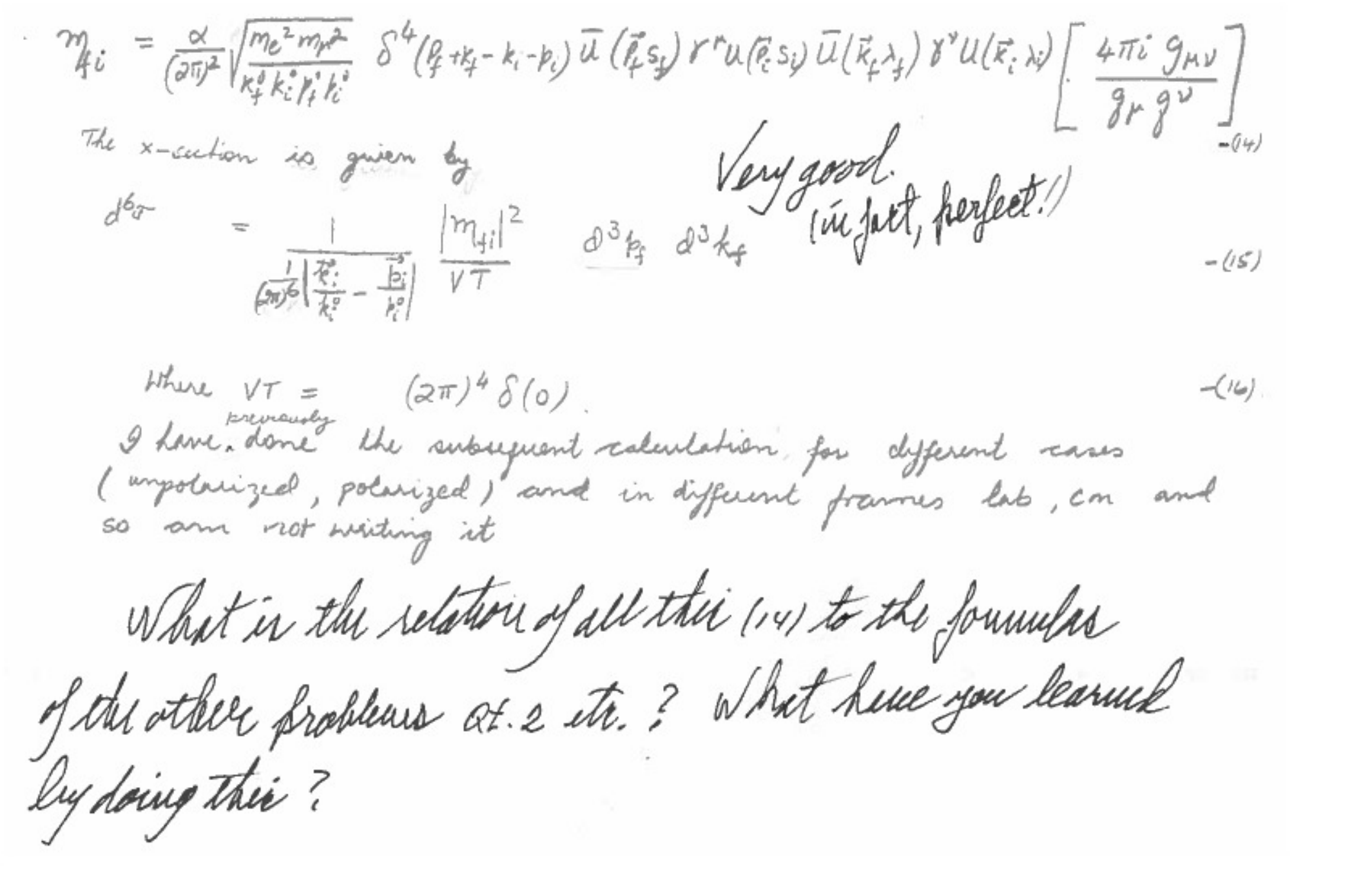}}
\caption*{\it Feynman's comments on Rajan Gupta's homework assignment}
\end{center}
\end{figure}

\newpage
\noindent\fbox{\parbox{0.98\linewidth}{
\begin{center}
{\large\bf Reminiscences of my Interactions with Feynman}
\end{center}

As a beginning graduate student, I was part of Feynman's study group.
The group included my batchmates Arun Gupta and Olivier Espinoza (now deceased),
some senior students such as Alexios Polychronakos (who left at some point after
graduation), and some undergraduates like Vineer Bhansali.

\hspace{0.5truecm}
Feynman's idea was to use the progress in integrability in two dimensions to study fragmentation or hadronisation in QCD. Consider a high energy scattering process, which gives rise to a quark or gluon moving at high energy that fragments as it moves away from the interaction region. There is one preferred direction along which the quark or gluon moves, where the momentum is high and the physics is `hard', and two transverse directions where it is `soft'. Feynman's idea was to separate these directions. Since high energy physics is simple in QCD due to asymptotic freedom, the hard physics would be easy to deal with. That would then leave the two transverse directions where the physics is soft and non-trivial, and perhaps the progress in integrability could be used here to understand the process of fragmentation.

\hspace{0.5truecm}
The first, and perhaps the most important lesson, for a beginning graduate student like me was just how thorough Feynman was. He believed that to understand what had been achieved in the study of integrability, we should simply solve all known integrable models! The Bethe ansatz, if I recall correctly, worked well, but only for some models. Feynman had already worked out these cases by the time I joined, and had detailed meticulous notes, a model for how careful and thorough one should be.

\hspace{0.5truecm}
However, and this is where things stood when I started interacting with him, the Bethe ansatz technique did not directly work for the Baxter 6 and 8 vertex models. Feynman explained the 6 vertex model to us, and his instructions were for us to try and solve it without looking at Baxter's solution, already in print. ``I have told you all you need to know," he said.
We met week after week on Wednesday afternoons and discussed the problem. We students made very little progress. Amongst the three of us, Arun made some headway, but we were still far from solving the model. It did not help that Baxter had been awarded the Heinemann prize that year for solving the 6 and 8 vertex models!

\hspace{0.5truecm}
Feynman was remarkably patient and spent a lot of time explaining the solutions that he had figured out using the Bethe ansatz technique for various models (which were all special cases of Baxter's models). We also went over a review on integrable models written by Hank Thacker, who Feynman always called ``Thaxter" (he also refered to the Zamolodchikov brothers as Zamaldovich and Zamaldovich driving us to splits of laughter). The high point for me of these sessions though were his digressions where he explained all kinds of physics to us.
Occasionally, the discussions would extend over dinner---for which Feynman often generously paid. I remember one great dinner at a Mexican restaurant on Colorado Boulevard, where he explained the spin-statistics theorem, using a belt which he took off from his pants, while exclaiming loudly to the waiter that she need not worry for he wasn't going to disrobe any further! This belt trick of Feynman is of course well-known and also discussed in his Dirac lecture.

\hspace{0.5truecm}
To come back to the Baxter models, one day Feynman called us to his office. His eyes were shinning. ``I have solved it!" he exclaimed. ``Get me some multi-colored chalk," he yelled to his secretary Helen. What followed was pure Feynman genius, in equal measure physics and showmanship of the highest order. The key, as Feynman explained, was to look at the lattice not in the standard fashion, as we had been doing, but at 45 degrees. Then due to the presence of particle number conservation in the 6 vertex case, a solution easily presented itself in a few lines of algebra, and the elliptic functions of Baxter's solution appeared almost miraculously! The underlying reason for the simplification at 45 degrees was that the system had effectively massless degrees of freedom. This was pure magic for me! Later on, experts told me that Feynman's solution is probably closely related to a well-known technique in integrable models involving corner transfer matrices. I have not followed up on this, but I am quite sure that Feynman did not know about this beforehand.

\begin{wrapfigure}{l}{0.35\textwidth}
\includegraphics[width=1.0\linewidth]{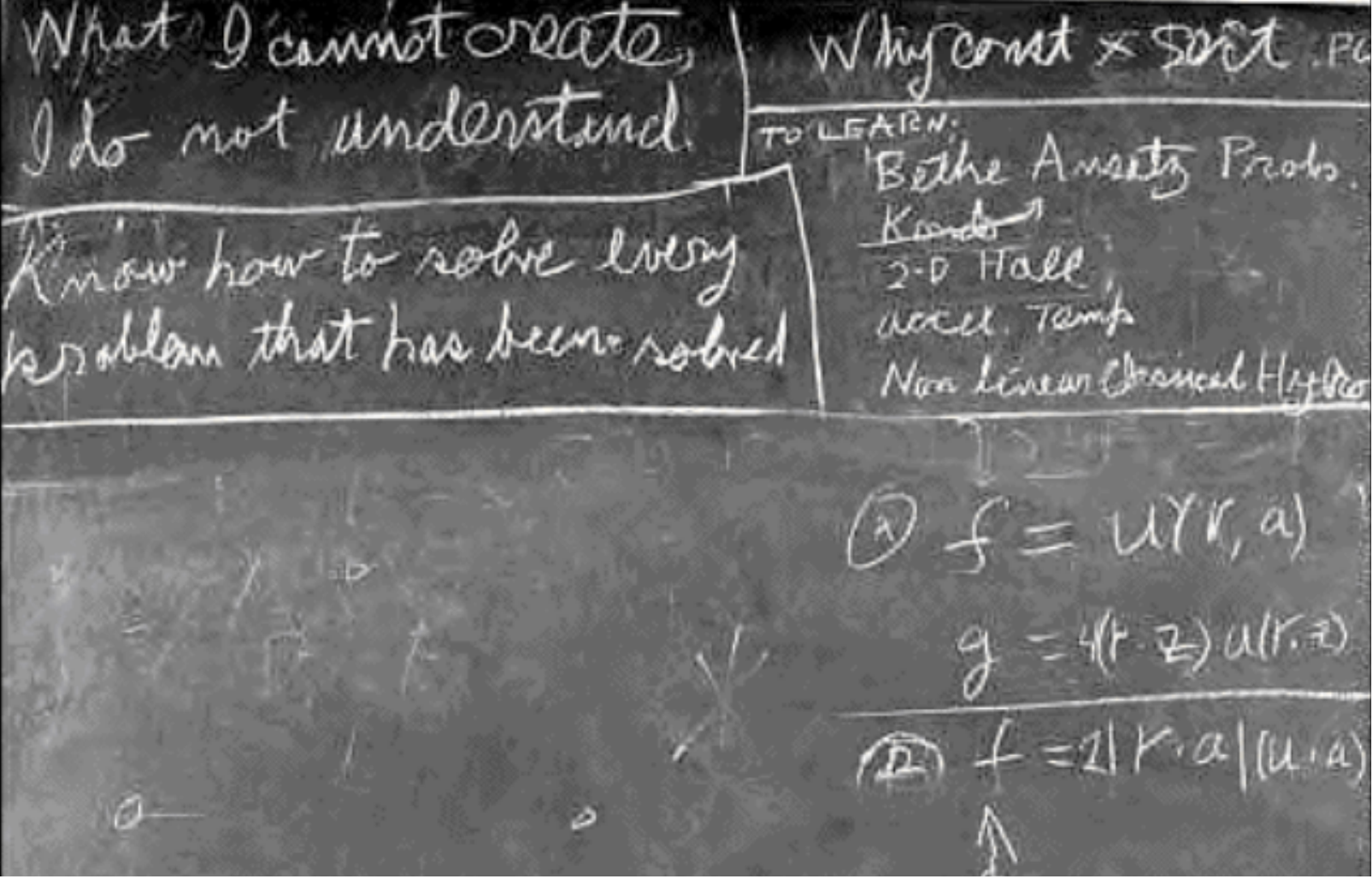} 
\caption*{\it{Feynman's last blackboard}}
\end{wrapfigure} 

\hspace{0.5truecm}
We students were then sent off to solve the 8 vertex model. This was more complicated because there was no particle number conservation, and particles could be created and destroyed. But Feynman was optimistic as ever. ``I invented all the tricks on how to deal with the creation and destruction of particles in QED," he said to us with a wink, ``We will surely get there." Sadly, Feynman fell quite ill soon thereafter, and our meetings grew infrequent. We never did solve the 8 vertex model, although we did sneak a look at Baxter's work, and figured out the solution that way. Feynman's dream of using integrability for QCD remained incomplete in our hands, and he passed away a few months later.

{\sl\hfill ---Sandip P. Trivedi}}}

\newpage
\noindent\fbox{\parbox{0.98\linewidth}{
\begin{center}
{\large\bf Richard P. Feynman's speech at the Nobel Banquet, 10 December 1965}
\end{center}

\begin{wrapfigure}{l}{0.25\textwidth}
\includegraphics[width=1.0\linewidth]{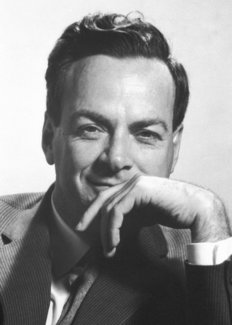} 
\end{wrapfigure} 

Your Majesty, Your Royal Highnesses, Ladies and Gentlemen.

The work I have done has, already, been adequately rewarded and recognized.
 
\hspace{0.5truecm}
Imagination reaches out repeatedly trying to achieve some higher level of understanding, until suddenly I find myself momentarily alone before one new corner of nature's pattern of beauty and true majesty revealed. That was my reward.
 
\hspace{0.5truecm}
Then, having fashioned tools to make access easier to the new level, I see these tools used by other men straining their imaginations against further mysteries beyond. There, are my votes of recognition.
 
\hspace{0.5truecm}
Then comes the prize, and a deluge of messages. Reports of fathers turning excitedly with newspapers in hand to wives; of daughters running up and down the apartment house ringing neighbors' doorbells with news; victorious cries of ``I told you so" by those having no technical knowledge---their successful prediction being based on faith alone; from friends, from relatives, from students, from former teachers, from scientific colleagues, from total strangers; formal commendations, silly jokes, parties, presents; a multitude of messages in a multitude of forms.
 
\hspace{0.5truecm}
But, in each I saw the same two common elements. I saw in each, joy; and I saw affection (you see, whatever modesty I may have had has been completely swept away in recent days).
 
\hspace{0.5truecm}
The prize was a signal to permit them to express, and me to learn about, their feelings. Each joy, though transient thrill, repeated in so many places amounts to a considerable sum of human happiness. And, each note of affection released thus one upon another has permitted me to realize a depth of love for my friends and acquaintances, which I had never felt so poignantly before.
 
\hspace{0.5truecm}
For this, I thank Alfred Nobel and the many who worked so hard to carry out his wishes in this particular way.
 
\hspace{0.5truecm}
And so, you Swedish people, with your honors, and your trumpets, and your king---forgive me. For I understand at last---such things provide entrance to the heart. Used by a wise and peaceful people they can generate good feeling, even love, among men, even in lands far beyond your own. For that lesson, I thank you. Tack!
}}

\end{document}